# Occupation and diffusion of interstitial solutes in dilute alloys in perspective of the Gauss–Legendre three-square theorem

Xiaoshuang Wang

Institute of Fluid Physics, China Academy of Engineering Physics, Mianyang 621999,China

Abstract:

In the example of the diffusion of C, N, O in dilute ferric iron alloys, it is shown that the polyhedron consisting of equivalent occupation of interstitial solutes in dilute alloys, can be classified into 7 groups altogether, i.e., cube, octahedron, cuboctahedron, truncated octahedron, truncated cube, rhombicuboctahedron and truncated cuboctahedron. No more polyhedron can be found. The notation and the migration paths of C, N, O in dilute alloys are well described by the Gauss–Legendre three-square theorem. The abstraction of the migration paths gives rise to the Wythoffian operations illustrated with cube and octahedron. The occupation and migration paths of the diffuser in this work can be generalized to the diffusion of vacancy and interstitial atoms in other host materials with low concentration of substitutional solutes.

⋆ corresponding author:

Xiaoshuang Wang

Institute of Fluid Physics

China Academy of Engineering Physics

Mianyang 621999

China

E-mail addresses: wangxshg@163.com

## 1. Introduction

Functional materials usually contain several foreign atoms, such as impurities during fabrication, or dopants and alloying elements in order to improve material properties. Foreign atoms or solutes with atomic sizes greater than or equal to those of host atoms migrate via the vacancy mechanism, and they mainly occupy substitutional sites. Solutes with atomic sizes smaller than those of host atoms diffuse via interstitial mechanism[1]. In this case, the most stable sites of solutes are highly symmetric interstitial sites in the lattice, i.e., octahedral and tetrahedral sites. When calculating the diffusivity of interstitial solutes, substitutional foreign atoms can be considered to be immobile, since the diffusivity of interstitial solutes is several orders of magnitude higher than that of substitutional foreign atoms. The migration of interstitial foreign atoms may be influenced by the presence of substitutional solutes, even if the concentration of substitutional solutes is below 0.1 at%. In dilute alloys, the concentration of substitutional solutes is so low that the diffusion of interstitial foreign atoms cannot be influenced simultaneously by more than one substitutional solute.

Substitutional sites for immobile solutes, lattice sites for host materials and interstitial sites for diffusing foreign atoms form the simple cubic (SC) lattice or a sublattice of the SC lattice. Present study takes diffusion of C, N, O in dilute ferric iron alloys for example. In body center cubic (BCC) Fe, the most stable sites of C, N, O are octahedral sites, which together with BCC lattice sites form SC lattice. The notation of octahedral interstitial sites in the neighborhood of a substitutional solute in a BCC lattice is a premise for this study. Barouh et al.[2, 3], Wang et al. [4-6] and Fedorov et al.[7] described these octahedral sites by neighboring sites (Fig. 1), where the *i*th neighbor site indicates the distance between interstitial sites and a substitutional solute is $\sqrt{i}$ times half of lattice parameter. Within the scheme of SC lattice, they claimed that 3$^{rd}$, 4$^{th}$, 7$^{th}$, 8$^{th}$, etc., neighbor positions cannot be occupied by interstitial solutes since these sites are already occupied by iron atoms. However, the 7$^{th}$ neighbor site is impossibly occupied by interstitial solutes or iron atoms, which is indeed the Gauss–Legendre three-square theorem.

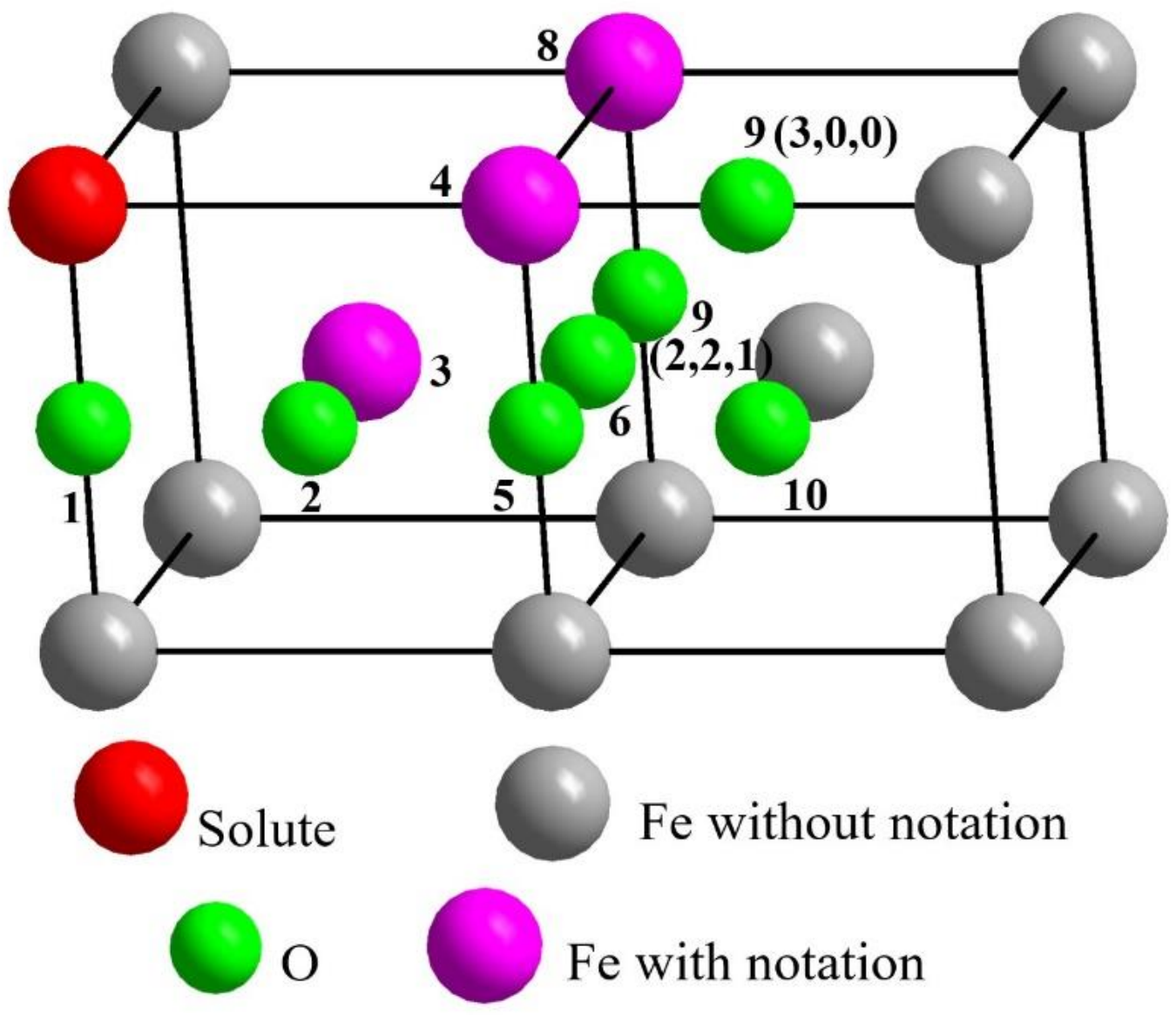

Fig. 1. Octahedral interstitial sites and lattice sites up to 10th neighbor shell relative to a substitutional solute.

The purpose of the present paper is to simplify the physical picture of atomic diffusion by means of the Gauss–Legendre three-square theorem. First, the Gauss–Legendre three-square theorem is applied to identify all the possible occupation sites of octahedral interstitial solutes under the influence of one substitutional solute in BCC lattice. Then, polyhedrons that use all equivalent *i*th neighbor sites of interstitial solutes as vertexes are constructed and classified. Finally, the relationship between these polyhedrons and migration paths of interstitial solutes in dilute alloys is discussed in detail.

2. the Gauss–Legendre three-square theorem

According to the Gauss–Legendre three-square theorem[8], any given positive integer $i$ is a sum of three squares precisely when $i$ is not of the form $4^k(8m+7)$ for any nonnegative integers $k$ and $m$. That is, if $i=a^2+b^2+c^2$ ($a$, $b$ and $c$ are nonnegative integers, $a \geq b \geq c$), then $i$ is not 7, 15, 23, 28, etc.. Therefore, the 7th, 15th, 23rd, 28th, etc., neighbor sites in the presence of one substitutional solute does not exist within the scheme of SC lattice. The maximal concentrations of substitutional solutes in order to include the octahedral interstitial sites within 7th, 15th, 23rd, 28th, neighbors are 1.85, 0.78, 0.40, and 0.23 at. %, respectively. The most attractive/repulsive interactions between C, N, O and substitutional solute usually occur at the 1st ,2nd or 5th neighbor distances. In the presence of substitutional solute, the most probable migration paths of C, N, O are 1st neighbor jumps from one modified octahedral site to another modified octahedral site. In the following only 1st neighbor jumps are considered. The jump between 5 and 10 is possible and is relevant to the possible influential 5th neighbor site. Therefore, the octahedral interstitial sites up to 10th neighbor shell should be a major consideration in the diffusion of interstitial solutes in dilute alloys. The octahedral interstitial sites beyond 10th neighbor and lattice sites are also investigated to get a more comprehensive picture of atomic diffusion.

3. Occupation and diffusion of interstitial solutes in dilute alloys

The octahedral interstitial sites and lattice sites up to 10th neighbor shell relative to the substitutional solute are shown in Fig. 1. Note that there are two different 9th neighbor sites, and ($a$, $b$, $c$) is an additional notation to distinguish the different octahedral interstitial and lattice sites with the same $i$. manipulating these neighbor sites gives new insight into the atomic diffusion. By connecting all equivalent 1st neighbor sites for interstitial solutes in the neighborhood of a substitutional solute, an octahedron is observed (Fig. 2(a)). Six vertexes of the octahedron are all equivalent 1st neighbor octahedral sites for interstitial solutes, and the substitutional solute sits in the geometric center of the octahedron. In the same way, a cuboctahedron (Fig. 2(b)) with twelve vertexes for all equivalent 2nd neighbor octahedral sites, a cube (Fig. 2(c)) with eight vertexes for all equivalent 3rd neighbor lattice sites, etc. forms. Table 1 lists the types of polyhedrons formed by all possible equivalent *i*th neighbor octahedral interstitial and lattice sites as vertexes with $i$ up to 20. The numbers with white, light gray and green backgrounds correspond to octahedral interstitial sites, lattice sites, and nonexistent sites, respectively. On the one hand, Table 1 demonstrates that $i$ up to 20 is not 7, 15. On the other hand, according to the polyhedral shape, there are only 7 types of polyhedrons altogether: cube, octahedron, cuboctahedron, truncated octahedron, truncated cube, rhombicuboctahedron and truncated cuboctahedron.

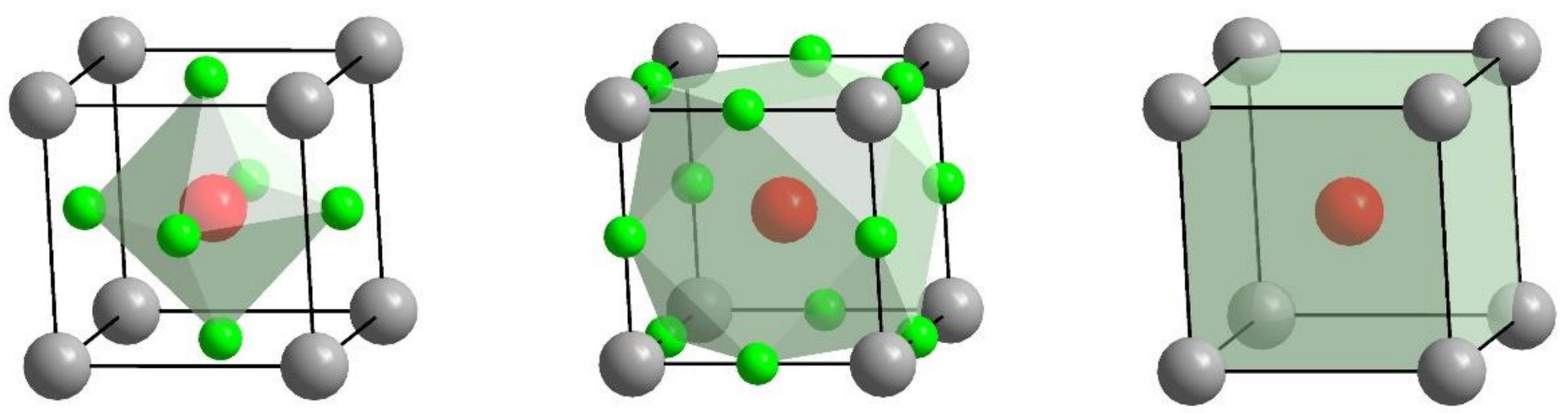

(a) $i$=1 (b) $i$=2 (c) $i$=3

Fig. 2. Polyhedrons consist of all equivalent 1st, 2nd neighbor interstitial sites and 3rd neighbor lattice sites.

Table 1. Types of polyhedrons consist of all equivalent $i$th neighbor octahedral interstitial and lattice sites as vertexes with $i$ up to 20.

| $i$ | ($a$, $b$, $c$) | Type |
|---|---|---|
| 1 | (1, 0, 0) | Octahedron |
| 2 | (1, 1, 0) | Cuboctahedron |
| 3 | (1, 1, 1) | Cube |
| 4 | (2, 0, 0) | Octahedron |
| 5 | (2, 1, 0) | Truncated octahedron |
| 6 | (2, 1, 1) | Rhombicuboctahedron |
| 7 | | |
| 8 | (2, 2, 0) | Cuboctahedron |
| 9 | (2, 2, 1) | Truncated cube |
| | (3, 0, 0) | Octahedron |
| 10 | (3, 1, 0) | Truncated octahedron |
| 11 | (3, 1, 1) | Rhombicuboctahedron |
| 12 | (2, 2, 2) | Cube |
| 13 | (3, 2, 0) | Truncated octahedron |
| 14 | (3, 2, 1) | Truncated cuboctahedron |
| 15 | | |
| 16 | (4, 0, 0) | Octahedron |
| 17 | (3, 2, 2) | Rhombicuboctahedron |
| | (4, 1, 0) | Truncated octahedron |
| 18 | (3, 3, 0) | Cuboctahedron |
| | (4, 1, 1) | Rhombicuboctahedron |
| 19 | (3, 3, 1) | Truncated cube |
| 20 | (4, 2, 0) | Truncated octahedron |

With increasing $i$ up to infinity, no more types of polyhedrons could be found. Based on $i=a^2+b^2+c^2$ ($a \geq b \geq c \geq 0$), the analysis of $a$, $b$ and $c$ leads out the following rule: 1) if $a=b=c\neq 0$, then the type of polyhedron with all possible $i$th neighbor octahedral interstitial or lattice sites as vertexes is cube ($a=b=c\neq 0$ ➔ cube); 2) $a\neq 0$, $b=c=0$ ➔ octahedron; 3) $a=b\neq 0$, $c=0$ ➔ cuboctahedron; 4) $a$, $b\neq 0$, $c=0$, $a\neq b$ ➔ truncated octahedron; 5) $a$, $b$, $c\neq 0$, $a=b>c$ ➔ truncated cube; 6) $a$, $b$, $c\neq 0$, $a>b=c$ ➔ rhombicuboctahedron; 7) $a$, $b$, $c\neq 0$, $a>b>c$ ➔ truncated cuboctahedron. It's worth mentioning that using $i=a^2+b^2+c^2$ implies that substitutional solute resides on the origin coordinates. The types, shapes and criterion, as well as the number of vertexes ($N_v$), faces ($N_f$) and edges ($N_e$) for all these 7 polyhedrons are summarized in Table 2, which is in consistent with the observations in Fig.2 and Table 1. the number of vertexes ($N_v$) is all possible equivalent *i*th neighbor octahedral interstitial or lattice sites. According to *a*, *b*, *c*, polyhedral shape could be imagined in the mind. Taken *a*, *b*, $c \neq 0$, $a > b = c$ as an example, the coordinates for the vertexes of the polyhedron are ($\pm a$, $\pm b$, $\pm b$), ($\pm b$, $\pm a$, $\pm b$) and ($\pm b$, $\pm b$, $\pm a$). The larger coordinates ±a determine 6 faces on which all these 24 vertexes reside, and ($\pm b$, $\pm b$) indicates that the polygon on each face among 6 is square with edges parallel to coordinate axis. Connecting these squares on 6 faces by line segments, it is exactly the shape of rhombicuboctahedron as shown in Table 2. The classification above covers all possibilities of *a*, *b*, *c*, which verifies the observation that only 7 types of polyhedrons could be found for any given *i*.

Table 2. Types, shapes and criterion, as well as the number of vertexes, faces and edges for all types of polyhedrons.

| Type | Shape | Criterion ($i=a^2+b^2+c^2$) | ($N_v$, $N_f$, $N_e$) |
|---|---|---|---|
| Cube | 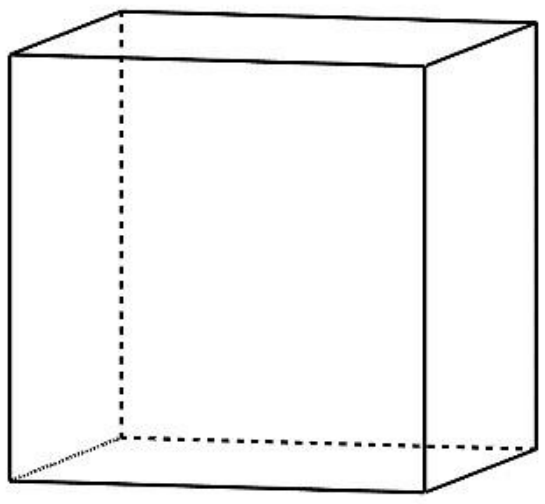 | $a=b=c\neq 0$ | (8, 6, 12) |
| Octahedron | 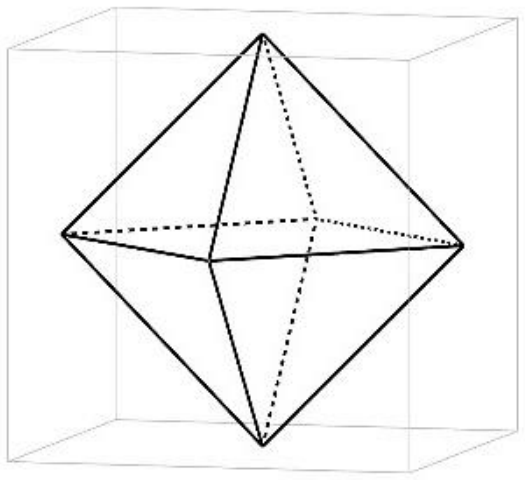 | $a\neq 0$, $b=c=0$ | (6, 8, 12) |
| Cuboctahedron | 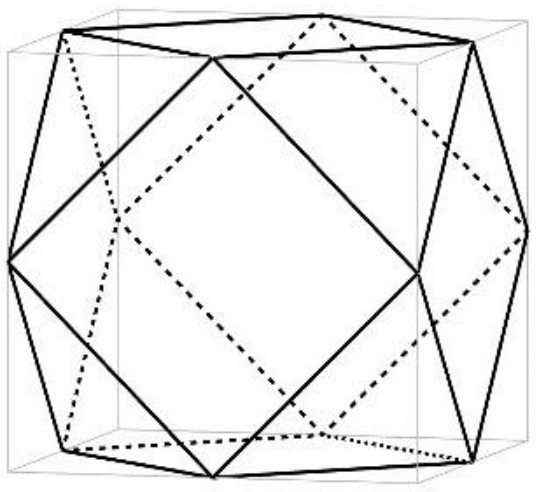 | $a=b\neq 0$, $c=0$ | (12, 14, 24) |
| Truncated octahedron | 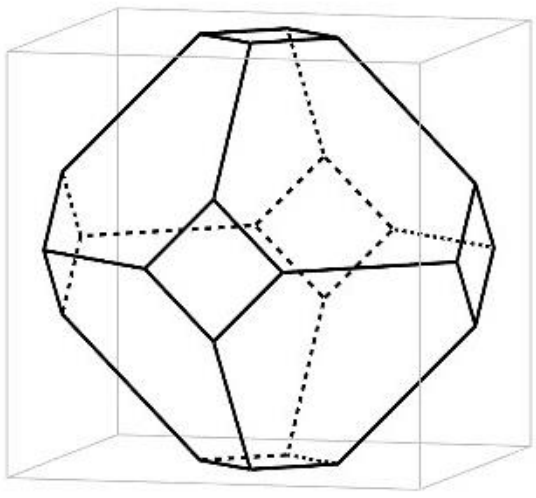 | $a$, $b\neq0$, $c=0$<br>$a\neq b$ | (24, 14, 36) |
| Truncated cube | 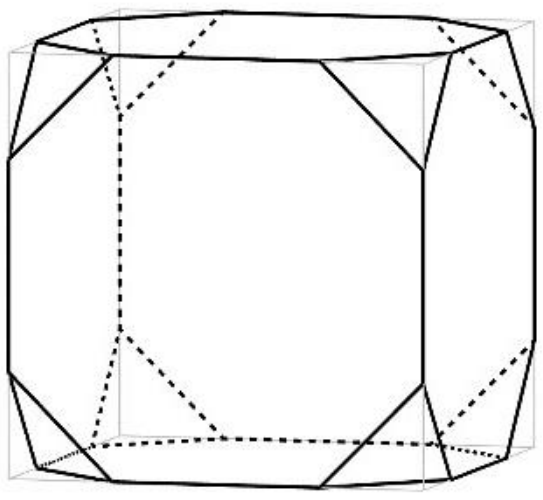 | $a$, $b$, $c\neq0$<br>$a=b>c$ | (24, 14, 36) |
| Rhombicuboctahedron | 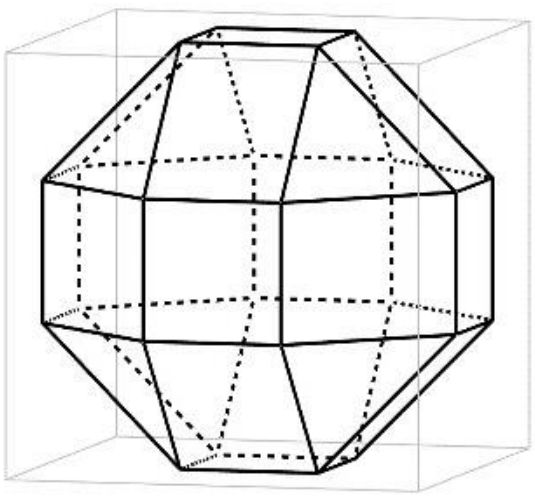 | $a$, $b$, $c\neq0$<br>$a>b=c$ | (24, 26, 48) |
| Truncated cuboctahedron | 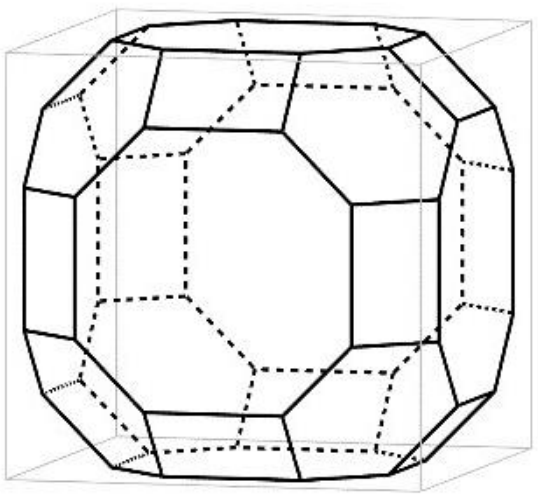 | $a$, $b$, $c\neq0$<br>$a>b>c$ | (48, 26, 72) |

Migration paths are generally prerequisites in calculations of diffusion coefficient of interstitial solutes in dilute alloys. The notation of ($a$, $b$, $c$) could be used to determine all probable migration paths as the 1st neighbor jump, since within the scheme for a SC lattice that consists of the BCC lattice sites and the octahedral interstitial sites of the BCC lattice, the 1st neighbor jump is one step jump along the direction of any coordinate axis, i.e., $a\pm1$, $b\pm1$ or $c\pm1$. After placing the substitutional solute on the origin coordinates, if $a$, $b$ and $c$ are all odd or even, ($a$, $b$, $c$) are lattice sites, otherwise octahedral interstitial sites. (1, 1, 1), (2, 0, 0), (2, 2, 0), (2, 2, 2), (3, 1, 1), (3, 3, 1), (3, 3, 3), (4, 0, 0), (4, 2, 0), (4, 2, 2), (4, 4, 0), (4, 4, 2) and (4, 4, 4) are lattice sites with $a$, $b$ and $c$ up to 4. $a$, $b$ and $c$ for octahedral interstitial sites are two odd and one even, or two even and one odd. The number of migration paths as the 1st neighbor jumps ($a\pm1$, $b\pm1$ or $c\pm1$) has the potential to be 6, among which two lead to jump between octahedral interstitial sites and BCC lattice sites because of the transition between even and odd due to ±1. Exclude these two impossible paths, maximum number of migration paths as the 1st neighbor jumps for one interstitial solute in an octahedral site is 4. All possible migration paths as the 1st neighbor jump with $a$, $b$ and $c$ up to 4 are shown as solid lines in Fig.3, while the dash lines indicate the impossible paths between octahedral sites and BCC lattice sites.

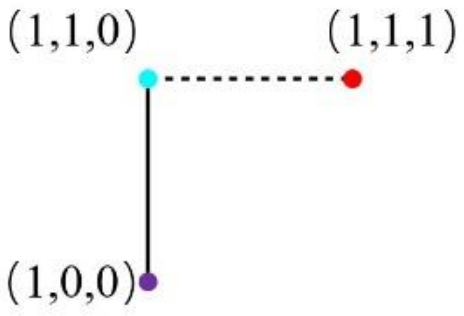

(a) 1st floor

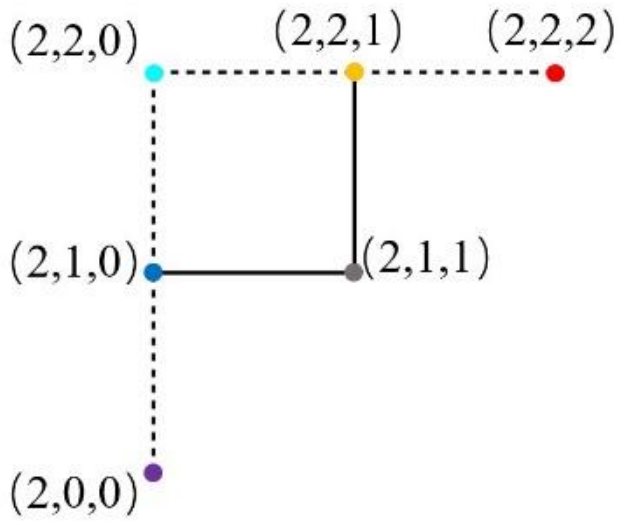

(b) 2nd floor

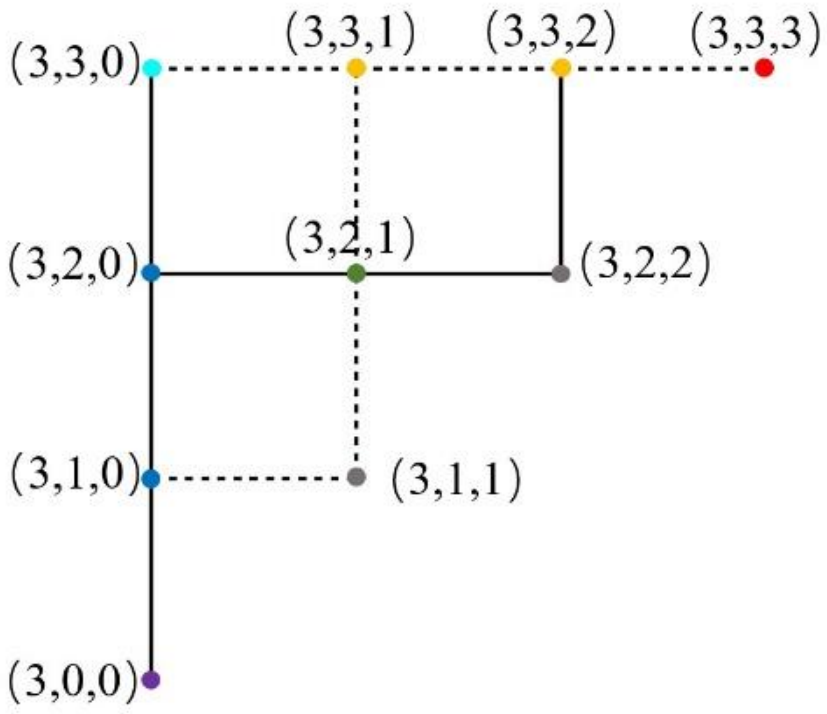

(c) 3rd floor

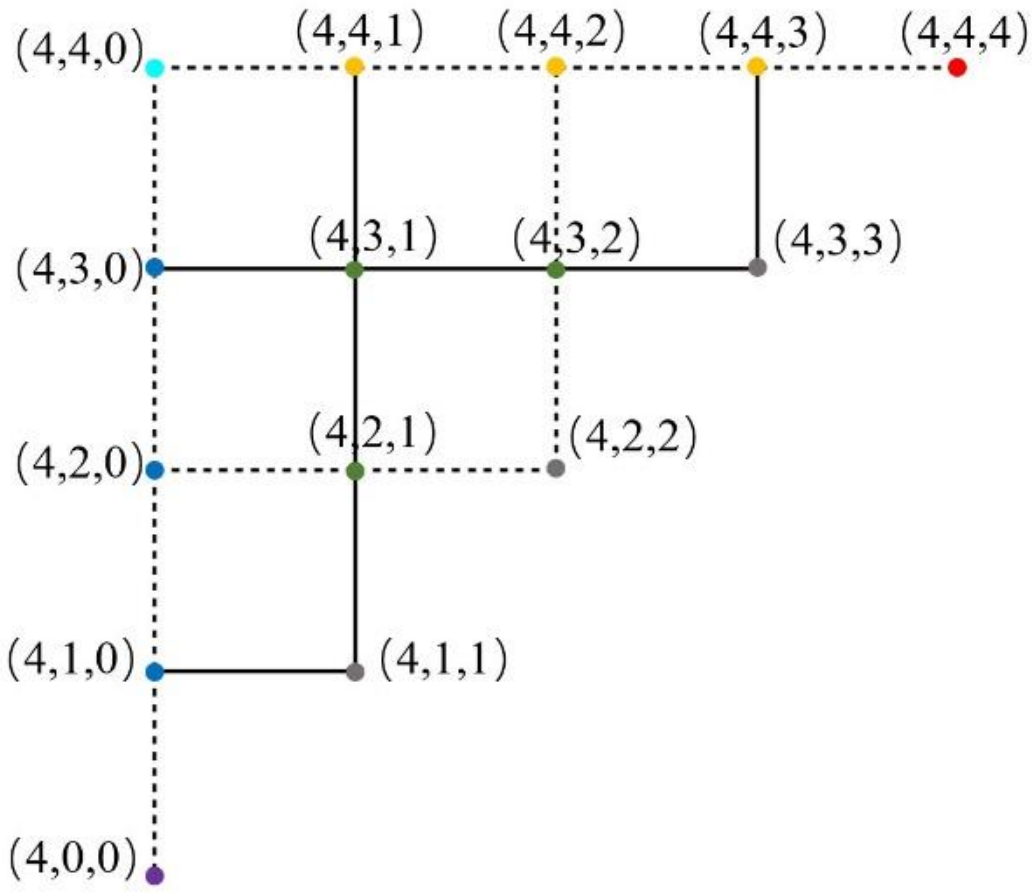

(d) 4th floor

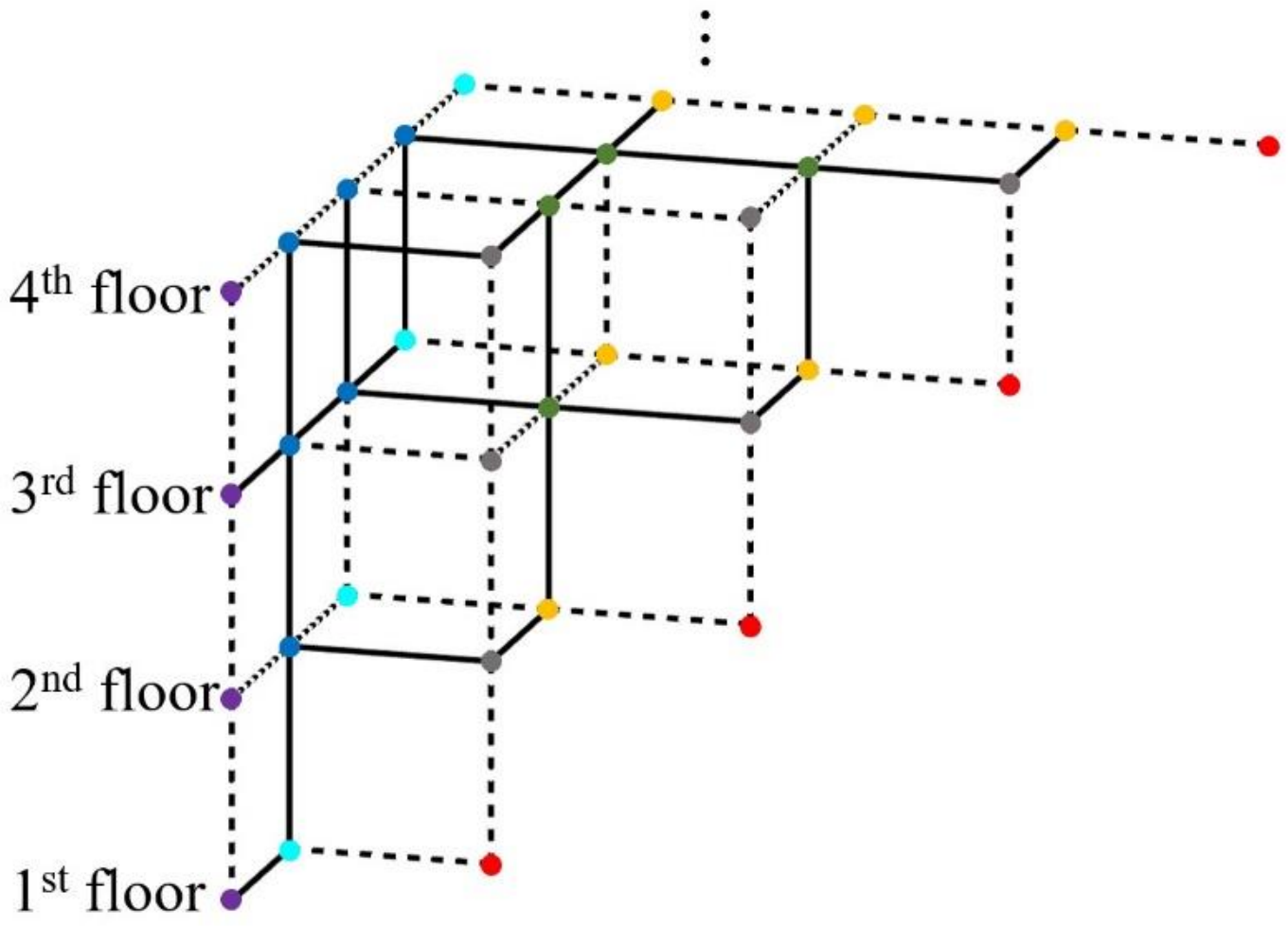

(e) Overall view

Fig. 3. All possible migration paths (solid lines) for C, N, O as the 1st neighbor jump in dilute ferric iron alloys with $a$, $b$ and $c$ up to 4. The red, violet, cyan, blue, golden, gray and green points correspond to cube, octahedron, cuboctahedron, truncated octahedron, truncated cube, rhombicuboctahedron and truncated cuboctahedron, respectively.

Abstracted from a top view of Fig. 3e, different types of polyhedrons that consist of equivalent

*i*th neighbor octahedral interstitial and lattice sites as vertexes are connected by migration paths as shown in Fig.4. Interestingly, Fig.4 is actually Wythoffian operations illustrated with cube and octahedron. The study of the Gauss–Legendre three-square theorem has a long history. The theorem was first proved by Legendre in 1798, and was studied and confirmed by various mathematicians in the next two centuries. But it is the first time that the Gauss–Legendre three-square theorem and Wythoffian operations for cube-octahedron family are related through the diffusion case. Fortunately, 7, the first absent number in the Gauss–Legendre three-square theorem, is back as the number of polyhedral shapes in Table 2 and Fig.4. Fig. 4 is also relevant to the typical polyhedral shape evolution of Ag nanocrystals using poly(vinylpyrrolidone) (PVP), i.e., cube → truncated cube → cuboctahedron → truncated octahedron → octahedron, which has been attributed to the stronger binding of PVP to Ag(100) facets than to Ag(111) facets.

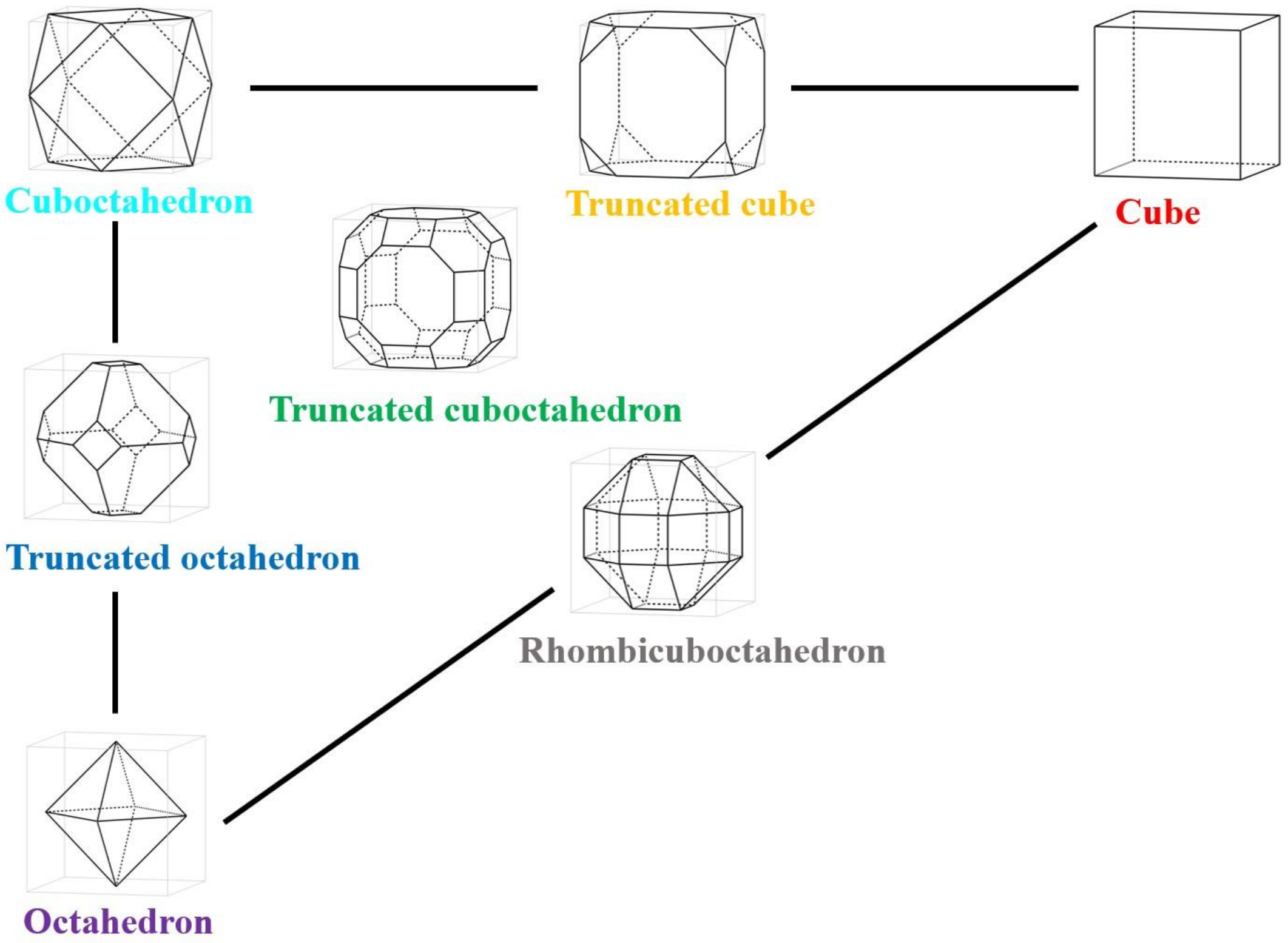

Fig. 4. Wythoffian operations illustrated with cube and octahedron from the abstraction of Fig. 3(e).

4. Conclusion

In conclusion, the octahedral interstitial sites and lattice sites in perspective of the Gauss–Legendre three-square theorem were investigated for the diffusion of C, N, O in dilute ferric iron alloys. According to the Gauss–Legendre three-square theorem, the notation of interstitial sites and lattice sites relative to the substitutional solute determine that the $7^{th}$, $15^{th}$, $23^{rd}$, $28^{th}$, etc., neighbor sites do not exist within the scheme of SC lattice. When connecting all possible equivalent *i*th neighbor octahedral interstitial or lattice sites as vertexes of a polyhedron, 7 types of polyhedrons could be found altogether, i.e., cube, octahedron, cuboctahedron, truncated octahedron, truncated cube, rhombicuboctahedron and truncated cuboctahedron. The notation of interstitial sites and lattice sites could be used to determine all probable migration paths of interstitial solutes as the $1^{st}$ neighbor jump. The Wythoffian operations illustrated with cube and octahedron is observed from the abstraction of the migration paths of C, N, O in dilute ferric iron alloys. The notation of the SC lattice that consists of octahedral interstitial sites and lattices sites and summary of possible migration paths as the 1st neighbor jump could be used for the diffusion of vacancy or other interstitial atoms in dilute alloys.

Acknowledgements
This work is supported by National Natural Science Foundation of China (Grant No.12205278 and 12141203).

References

[1] H. Mehrer, Diffusion in solids: fundamentals, methods, materials, diffusion-controlled processes, Springer Science & Business Media2007.
[2] C. Barouh, T. Schuler, C.-C. Fu, M. Nastar, Physical Review B, 90 (2014).
[3] C. Barouh, T. Schuler, C.-C. Fu, T. Jourdan, Physical Review B, 92 (2015).
[4] X. Wang, M. Posselt, J. Faßbender, Physical Review B, 98 (2018).
[5] X. Wang, J. Faßbender, M. Posselt, Physical Review B, 101 (2020).
[6] X. Wang, J. Fassbender, M. Posselt, Materials (Basel), 12 (2019).
[7] M. Fedorov, J.S. Wróbel, A.J. London, K.J. Kurzydłowski, C.-C. Fu, T. Tadić, S.L. Dudarev, D. Nguyen-Manh, Journal of Nuclear Materials, 587 (2023).
[8] P. Pollack, P.J.M.o.C. Schorn, 88 (2019) 1007-1019.